\documentclass[pra,aps,a4paper,twocolumn,superscriptaddress,nofootinbib]{revtex4-2}

\usepackage[T1]{fontenc}
\usepackage[utf8]{inputenc}
\usepackage{amssymb}
\usepackage{amsmath}
\usepackage{graphicx}
\usepackage{bbm}
\usepackage{psfrag}
\usepackage{hyperref}
\usepackage{braket}
\usepackage[normalem]{ulem}

\usepackage{xcolor}
\hypersetup{
    colorlinks=true,
    citecolor=blue,
    linkcolor=blue,
    filecolor=magenta,
    urlcolor=blue}
\definecolor{ccqqqq}{rgb}{1,0,0}
\definecolor{uuuuuu}{rgb}{0.26666666666666666,0.26666666666666666,0.26666666666666666}
\definecolor{qqwwzz}{rgb}{0,0,1}
\usepackage{float}

\newcommand{\beq}{\begin{equation}}
\newcommand{\eeq}{\end{equation}}
\newcommand{\bea}{\begin{eqnarray}}
\newcommand{\eea}{\end{eqnarray}}
\newcommand{\bit}{\begin{itemize}}
\newcommand{\eit}{\end{itemize}}

\def\Tr{{\rm Tr}}

\def\le{\left(}
\def\ri{\right)}

\usepackage{geometry}
\newgeometry{includefoot,left=2cm,right=2cm,bottom=1cm,top=2cm}

\newcommand{\eg}{{\it e.g.,}\ }
\newcommand{\ie}{{\it i.e.,}\ }

\makeatletter
\DeclareRobustCommand{\cev}[1]{%
  \mathpalette\do@cev{#1}%
}
\newcommand{\do@cev}[2]{%
  \fix@cev{#1}{+}%
  \reflectbox{$\m@th#1\vec{\reflectbox{$\fix@cev{#1}{-}\m@th#1#2\fix@cev{#1}{+}$}}$}%
  \fix@cev{#1}{-}%
}
\newcommand{\fix@cev}[2]{%
  \ifx#1\displaystyle
    \mkern#23mu
  \else
    \ifx#1\textstyle
      \mkern#23mu
    \else
      \ifx#1\scriptstyle
        \mkern#22mu
      \else
        \mkern#22mu
      \fi
    \fi
  \fi
}
\makeatother

\begin{document}

\author{Tal Schwartzman}\email{taljios@gmail.com\\ \\}

\affiliation{\it Department of Physics, Ben-Gurion University of the Negev, \\ David Ben Gurion Boulevard 1, Beer Sheva 84105, Israel}

\title{The complexity of entanglement embezzlement}

\begin{abstract}
Embezzlement of entanglement is the counterintuitive process in which entanglement is extracted from a resource system using local unitary operations, with almost no detectable change in the resource's state. It has recently been argued that any state of a relativistic quantum field theory can serve as a resource for perfect embezzlement. We study the circuit complexity of embezzlement, using sequences of states that enable arbitrary precision for the process, commonly called universal embezzling families. In addition, we argue that this approach provides a well-defined model for the complexity of embezzlement from quantum field theories. Under fairly general assumptions, we establish a generic lower bound on the complexity, which increases with the precision of the process or embezzled entanglement, and diverges as these become infinite. As an example, we consider a $1d$ critical system as the resource and derive an exponentially growing lower bound on the complexity. Consequently, the findings imply that circuit complexity acts as a physical obstruction to perfect embezzlement. Supplementary to the main results, we derive lower bounds for common models of circuit complexity for state preparation, based on the difference between the Schatten norms of the initial and final states. 
\end{abstract}

\maketitle

\section{Introduction}
Entanglement is a key ingredient of modern quantum technologies, serving as a valuable resource for information processing \cite{horodecki2009quantum, nielsen2010quantum}. Beyond its practical uses, entanglement also has indispensable theoretical value throughout modern physics, from condensed matter where it can be used to diagnose phases of matter, to quantum gravity where it helps explain the emergence of spacetime in holographic models, (see, \eg  \cite{amico2008entanglement, ryu2006holographic, van2010building, swingle2012entanglement}).
The defining feature of entanglement is that it cannot be increased by local operations and classical communication \cite{vidal2000entanglement}. It might come as a surprise then, that there is a process in which entanglement can be extracted using local unitary operations without almost any detectable change in the system from which it was extracted. This process is known as \textit{embezzlement of entanglement} \cite{van2003universal}.

Consider a partitioned system 
having a Hilbert space $\mathcal{H}_{A_c}\otimes \mathcal{H}_{A_e}\otimes \mathcal{H}_{B_c}\otimes \mathcal{H}_{B_e}$, where Alice has access to system $A$, and Bob to $B$. The system $\mathcal{H}_{A_e} \otimes \mathcal{H}_{B_e}$ will be referred to as the embezzling system, and the rest as the catalyst, from which entanglement is being extracted. 
The protocol of entanglement embezzlement from a catalyst state $\ket{\Omega_c} \in \mathcal{H}_{A_c}\otimes \mathcal{H}_{B_c}$, is done by acting with a local unitary $U_A \otimes U_B$, such that
\begin{equation} \label{embezzling}
    \big|\bra{\Omega_c, \psi_e} U_A \otimes U_B \ket{\Omega_c, \phi_{A_e} \phi_{B_e}}\big| >1- \frac{\varepsilon}{2}.
\end{equation}
where $\ket{\phi_{A_e} \phi_{B_e}}$ and $\ket{\psi_e}$ are states of the embezzling system, with $\ket{\psi_e}$ being an entangled state, and $\varepsilon$ is a small parameter. As local unitary operations cannot create entanglement, this protocol seems counterintuitive. In other words, there is an apparent contradictory information loss, as the von Neumann entropy of either Alice or Bob's reduced density matrix seems to increase by a unitary operator.

Nevertheless, van Dam and Hayden \cite{van2003universal} showed that for the state $\ket{\Omega_c} = C\sum_{j=1}^N \frac{1}{\sqrt{j}}\ket{jj}$ belonging to an $N^2$-dimensional Hilbert space, one can perform the protocol with $\varepsilon =  2 \frac{ \log{d_e}}{\log{N}}$, where $d_e^2$ is the Hilbert space dimension of the embezzling system. Therefore, for large enough $N$, embezzlement can be performed to good precision. On the other hand, for any $N$, there are embezzling systems with large enough $d$ such that $\varepsilon$ reaches its maximal value of $2$. Entanglement embezzlement has various applications in quantum information theory, including its role as an important ingredient in the quantum reverse Shannon theorem \cite{berta2011quantum, bennett2014quantum}, and in winning strategies of quantum non-local games \cite{leung2013coherent, dinur2015parallel, regev2015quantum, cleve2017perfect, ji2020three}. One might wonder whether there are states, at least in infinite dimensional systems, that can act as perfect catalysts for embezzlement. In \cite{cleve2017perfect}, the authors found that perfect embezzling is possible, but cannot be achieved for a catalyst Hilbert space with a tensor product structure between $A$ and $B$.

Recently, it was argued that relativistic quantum field theories are universal embezzlers: any entangled state of any dimension can be embezzled from them with arbitrary precision \cite{van2024embezzling, van2024embezzlement}.\footnote{More generally, this was shown for catalysts for which the $A$ and $B$ subsystem algebras of bounded operators are type III$_1$ factors.} For this catalyst system, there is no tensor product structure for the $A$ and $B$ subsystems, but instead, there is a commuting operator framework: the notion of locality is imposed by having the operations of Alice and Bob in their respective labs, such as $U_A$ and $U_B$, commute. To give an example, the catalyst state can be the vacuum state of a $1+1$ dimensional relativistic quantum field and $U_{A(B)}$ is generated by operators in the left(right) Rindler wedge that couple to the embezzling system in $A(B)$. In this example embezzling of entanglement is what is often denoted in the literature as \textit{entanglement harvesting} \cite{reznik2003entanglement, reznik2005violating, pozas2015harvesting, perche2024fully, bozanic2023correlation}, where the detectors are causally disconnected, but with a requirement on the final state of the field to be (almost) unperturbed.

However, the ability to perfectly embezzle from a quantum field presents a puzzle. Intuitively, embezzling a larger amount of entanglement, or to a higher degree of precision, will require $U_A$ and $U_B$ to probe regions closer to the boundary between $A$ and $B$, including higher energy modes. Therefore, for an actual implementation of the protocol, one can expect to encounter some physical obstruction rendering perfect embezzlement impossible. What, then, is this physical limitation? One candidate is the energy needed for the implantation of a unitary on the system \cite{tajima2018uncertainty, takagi2020universal, chiribella2021fundamental, yang2022energy}. Yet in the embezzlement protocol, better precision imposes tighter bounds on the distance between the final and initial states of the catalyst, which, in turn, constrains the change in energy induced by the protocol.\footnote{We note that this is not a rigorous statement. In finite-dimensional systems, one can upper bound the energy difference between two states that respect \eqref{embezzling}, by $|\Delta E| \leq \|H\| 2 \sqrt{ \varepsilon}$ where $\|*\|$ is the operator norm. However, finite-dimensional catalyst systems cannot accommodate an ever-decreasing $\varepsilon$. It would be interesting to further investigate how the minimal possible energy difference, induced by a protocol with a physical catalyst, scales with the precision of the protocol.} Therefore it is not obvious that the energy needed will serve as an obstacle. In addition, as the protocol is purely kinematic, \ie the Hamiltonians of the systems do not participate in it, there should be a kinematic obstruction, independent of the details of the embezzling system, such as the energy difference between its initial and final states, $\ket{\phi_{A_e} \phi_{B_e}}$ and $\ket{\psi}$. The issue is resolved for finite-dimensional systems, as the embezzling capabilities \ie the best possible $\varepsilon$ are fixed with the catalyst's size and decrease with the amount of embezzled entanglement.

In this manuscript, we show that indeed there is an intuitive physical obstruction to the embezzlement protocol, which is given by the circuit complexity of the unitaries implementing it. Quantum circuit complexity quantifies the difficulty of performing certain tasks and in general, aims to characterize the capabilities of quantum computers and their advantage over classical ones (see, \textit{e.g.}, \cite{bernstein1993quantum, Watrous2009}). Independently of quantum computation, in the last decade complexity has received much attention as a quantum informational quantity with wide applications, ranging from the study of chaos \cite{ali2020chaos, balasubramanian2020quantum, Balasubramanian:2021mxo, craps2022bounds, qu2022chaos, craps2024integrability}, to topological phase transitions \cite{liu2020circuit} and having a vast interest in high energy physics, specifically in the realm of holography  \cite{Susskind:2014rva, Brown:2015bva, couch2017noether, Caputa:2018kdj, Erdmenger:2021wzc, Belin:2021bga, Chapman:2021eyy, Baiguera:2023tpt, Aguilar-Gutierrez:2023zqm}.\footnote{see \cite{Chapman:2021jbh} for a review on complexity in holography.}

We study lower bounds on the complexity of embezzlement for general \emph{universal embezzling families} \cite{leung2014characteristics, zanoni2024complete}, which are sequences of states (similar to those in \cite{van2003universal}) that allow for arbitrary embezzlement precision. Our results establish a lower bound on the complexity of embezzlement, applicable to any such family. This bound grows polynomially with the embezzlement precision or the amount of entanglement being embezzled and diverges in the limit of perfect precision or infinite entanglement. Circuit complexity depends on how one classifies “easy” versus “hard” gates. The scaling of our bound with the precision and entanglement is valid for a large class of commonly used circuit models, consisting of gates limited in their range of interaction.

In addition, we analyze the scenario where the embezzling family arises from a $1d$ critical system. In this case, we obtain a lower bound that increases exponentially with both the precision and the embezzled entanglement. To achieve this, we derive a novel lower bound on the circuit complexity of state preparation, in terms of the change of the state's Schatten norms, a result that may be of independent interest (see Appendix \ref{SchattenNormBound}).

Beyond our primary goal, these findings contribute to the broader study of complexity itself. The states we investigate can be viewed as regularized quantum field states, where the regularization cutoff grows with the embezzlement precision. Defining complexity in field theories or continuous-variable systems remains an active research area \cite{Chapman:2017rqy, Jefferson:2017sdb, Flory:2020dja, Flory:2020eot, Chagnet:2021uvi}. Our results suggest a lower bound on the complexity of a protocol carried out in a quantum field theory that does not depend on the regularization cutoff.

\section{Circuit complexity}
Quantum circuit complexity of a unitary $U$ counts the minimal number of gates needed out of a universal set, to create $U$. If certain gates are harder than others, extra weight will be given to them in the counting. Nielsen showed that lower bounds on quantum circuit complexity can be achieved by minimizing the length of a trajectory in the special unitary group that connects the identity and $U$, where the manifold is equipped with a suitable norm on the tangent space \cite{Nielsen1,Nielsen_2006, nielsen2006optimal, Nielsen3}. This length is often considered a unitary complexity measure in its own right.

A circuit that acts on an $N$-dimensional Hilbert space can be thought of as a parameterized unitary $U(t) \in$ SU($N$), with
\begin{equation}
    U(t) = \mathcal{\cev{P}} \exp \le -i \int_0^t dt' H(t') \ri \, ,
\label{eq:generic_path_unitaries}
\end{equation}
with $U(0) = \boldsymbol{1}$ and $U(1) = U$, where $\mathcal{\cev{P}}$ denotes the path ordering such that $\dot{U}(t) = -i H(t) U(t)$. $H(t)$, the control Hamiltonian, is a traceless Hermitian operator that belongs to the algebra of the SU($N$) generators. The circuit cost is a functional of $U(t)$, which "counts" the gates in the circuit with the specific choice of weights. As the infinitesimal gate that takes $U(t)$ to $U(t+dt)$ is $e^{-i dt H(t)}$, we can consider functionals of $H(t)$. Suppose that it is only possible to generate evolution with a certain set of infinitesimal gates, such that $H(t)$ is restricted to have the form $H(t) = \sum_I Y_I(t) T_I$, where $T_I$ is an element of a subset of the traceless Hermitian generators of SU($N$). 
We shall focus on a cost functional of the form,\footnote{Nielsen considered a norm on $H$ that penalized hard gates such that in the limit of large penalty they will approximately not be used \cite{Nielsen1,Nielsen_2006, nielsen2006optimal, Nielsen3}. Here we assume a restricted Hamiltonian from the start.}
\begin{equation}
   \text{cost}(U(t)) =  \int_0^1 dt \, \sum_I |Y_I (t)| \, .
\label{eq:cost} 
\end{equation}
The unitary complexity is the minimal cost among all such circuits that realize $U$,
\begin{equation} 
    C(U) = \min_{\lbrace Y^I : \, U(0)=\mathbf{1}, \, U(1)=U \rbrace} \text{cost}(U(t))\, .
\label{eq:unitary_complexity} 
\end{equation}
If the set of possible gates is discrete and of the form of $\{e^{-i T_I}\}$, the cost of a circuit composed of them will be exactly the number of gates and the above continuous version will give a lower bound.

\section{The complexity of embezzlement}
Instead of the bi-partite picture, we simplify the analysis by considering the protocol reduced to one of the parts. The authors of \cite{van2024embezzlement} showed that, if there is a unitary $U_A$, such that  
\begin{equation} \label{onesideEmbezz}
\|U_A \omega \otimes \phi_{A_e} U_A^\dagger-\omega \otimes \psi_{A_e}\|_1 \leq \varepsilon
\end{equation}
then there is a unitary $U_B$ for which \eqref{embezzling} holds. Here, $\|*\|_1$ is the trace norm,\footnote{More carefully, let $X_A$ be the algebra of bounded operators to which Alice has access, equipped with the operator norm $\|*\|$. We define the norm $\|\xi\|_1$ of a linear functional $\xi$ which assigns expectation values to operators in $X_A$, as $\|\xi\|_1 = \max_{\|X_A\|=1} |\xi(X_A)|$. In the case that $\xi$ can be represented by a finite-dimensional matrix, this is exactly the trace norm.}  $\phi_{A_e}$ and $\psi_{A_e}$ are the reduced density matrices of $\ket{\phi_{A_e} \phi_{B_e}}$ and $\ket{\psi_e}$ respectively, and $\omega$ is the quantum state on $X_{A_c}$ which is the algebra of bounded operators on the part of the catalyst to which Alice has access, defined as $\omega(X_{A_c}) = \bra{\Omega_c} X_{A_c} \ket{\Omega_c}$.\footnote{If the Hilbert space of the catalyst factorizes, the trace distance between two states, which is half the trace norm of their difference, is lower bounded by $1-\sqrt{F}$, where $F$ is the fidelity \cite{fuchs1999cryptographic}. Thus, if \eqref{onesideEmbezz} holds, then $1-\sqrt{F(U_A \omega \otimes \phi_{A_e} U_A^\dagger,\omega \otimes \psi_{A_e})} \leq \frac{\varepsilon}{2}$. Uhlmann's theorem \cite{uhlmann1976transition} then implies that there is a unitary $U_B$ such that \eqref{embezzling} also holds.} Our goal is to consider the protocol to any precision, regardless of the embezzled state. Embezzling states, which are states that allow exactly that, cannot be represented by a reduced density matrix \cite{van2024embezzling, van2024embezzlement}, and therefore we avoid referring to $\omega$ as such. Assuming Alice and Bob build their unitary circuits from gates local to their respective labs, the complexity of $U_A\otimes U_B$ in \eqref{embezzling} will simply be the sum of the complexities of $U_A$ and $U_B$.
Therefore, from here on, we shall consider the complexity of performing the one-sided protocol and drop the subscript denoting the subsystem. The catalyst and embezzling systems will now denote what was previously their one-sided parts, \ie the parts to which Alice has access.

In realistic scenarios, practical control over a quantum system, as well as the resolution of measurement outcomes, are often digitalized or coarse-grained. This means that the state can effectively be described by a density matrix on a Hilbert space. Furthermore, circuit complexity is most naturally formulated for finite-dimensional systems, where it is well-understood over the entire space of possible unitaries. 
Therefore, instead of $\omega$, we consider an embezzling family of states $\omega_n$, each acting on the Hilbert space $\mathcal{H}_n$ of $n$ qudit sites with local dimension $d$,\footnote{The statements of Sec. \ref{criticalSec}, however, do not depend on $d$, which can be taken to infinity.} and we remind that the dimension of the embezzling system is $d_e$, \ie 
\begin{equation}
\begin{split}
\omega_n \;\in\; & \mathcal{D}(\mathcal{H}_n),
\quad
\dim(\mathcal{H}_n) \;=\; d^n,
\\
\phi,  \psi \;\in\; & \mathcal{D}(\mathcal{H}_e),
\quad
\dim(\mathcal{H}_e )\;=\; d_e,
\end{split}
\end{equation}
where $\mathcal{D}(\mathcal{H}_n)$ is the space of positive semi-definite, self-adjoint operators of trace one acting on the Hilbert space. In the limit of $n \to \infty$, there is a unitary such that finite entanglement can be embezzled from $\omega_n$, with $\varepsilon \to 0$. 

If one has a specific embezzling state $\omega$ in mind, $\omega_n$ can be seen as either a coarse-grained, partially traced, or a regularized version of it, capturing more of its information as $n$ increases (see \cite{van2024multipartite} for a discussion on relations between embezzling states and embezzling families).  
In relativistic quantum field theory, algebras of local observables are of type III \cite{driessler1977type, longo1982algebraic, Buchholz1987, yngvason2005role}. In this context, $\omega_n$ can be understood as a truncation of $\omega$ to a type I subfactor. For example the state on $n$ commuting smeared field operators with an appropriate regularization that makes the dimension finite, or the state of a subregion in a field theory lattice discretization scheme in which $n$ controls the lattice spacing. In addition, $\omega_n$ arises naturally in the construction of type III algebras, as infinite tensor products of finite type I factors \cite{araki1968classification, blackadar2006operator, witten2018aps} (see the section about infinite tensor products in \cite{van2024embezzlement} for a detailed example of $\omega_n$) and in certain qudit chains in the thermodynamic limit \cite{van2024critical}.
We therefore define,
\begin{equation} \label{finiteCompOfEmb}
    \begin{aligned}
        C_n(\varepsilon) = \min_{U} C(U) \ \ \text{s.t} \ \ \| U \omega_n \otimes \phi U^\dagger - \omega_n \otimes  \psi\|_1 < \varepsilon.
    \end{aligned}
\end{equation}

For any finite $n$, $\varepsilon$ in \eqref{finiteCompOfEmb} can take values that are strictly larger than $0$ (assuming $\psi$ and $\phi$ have a different spectrum). In addition, it is possible that for a fixed value of $\varepsilon$, a larger number of sites, \ie a larger $n$, allows for a lower complexity.  We therefore define the complexity of embezzlement as
\begin{equation}\label{embezzlingComp}
    \begin{aligned}
        C(\varepsilon) = \min_n C_n(\varepsilon).
    \end{aligned}
\end{equation}

Let us now specify the cost function on the unitary circuits. From physical constraints, it is common to consider only $k$-local interactions in the possible gates of the circuit. Here we shall consider $k$ - geometrically local interactions, \ie the allowed terms in the Hamiltonian will couple at most $k$ nearest neighbor sites. We leave the general analysis of the non-geometrically local case for future studies. See, however, Sec. \ref{Other} where we consider it with specific types of cost functionals.
We start with $k=2$. Let us restrict the circuit's Hamiltonian to
\begin{equation} \label{controlH}
    H = \sum_{I}Y_{I} T_I = \sum_{I=1} h_I
\end{equation}
where the $T_I$ operators are 2-local traceless hermitian operators having unit operator norm with support on the different nearest neighboring sites, and $h_I =Y_{I} T_I$. For example if $d=2$, a common set of generators is given by the tensor product of Pauli matrices, $h_I = Y^i_{a,\mu}\sigma^i_a \otimes \sigma^{i+1}_\mu$ where different $I$ indices correspond to different choices of $a \in [1,3] ,\mu \in [0,3]$ and $i$, with $i$ marking the site number and $a$ and $\mu$ mark the type of Pauli matrix.  
The cost of this circuit is given by
\begin{equation} \label{costNorm}
    \text{cost}(U(t)) = \int_0^1 dt \sum_{I}| Y_{I}(t)| =  \int_0^1 dt\sum_{I}\|h_I(t)\|,
\end{equation}
where $\|*\|$ is the operator norm.

Let us now introduce a convenient notation. As explained above \eqref{finiteCompOfEmb}, the Hilbert space to which $\omega_n$ belongs, is of a chain of $n$ sites with local dimension $d$, and the Hilbert space of the (one-sided) embezzling system is of a site with dimension $d_e$.
let us set the embezzling system to site $0$, and denote by
\begin{equation}\label{rhoiDef}
    \rho_i^{(n)}(t) \equiv \Tr_{i+1,...,n}U(t)\omega_n \otimes \phi U(t)^\dagger
\end{equation} 
the reduced density matrix of the first $i$ sites of the catalyst, plus the embezzling system, time evolved along the circuit trajectory that gives the minimal cost. Notice that $U(0)$ is the identity, and $U(1)$ is the entire circuit that realizes the embezzlement protocol, and therefore $\rho_i^{(n)}(0) \equiv \Tr_{i+1,...,n}\omega_n \otimes \phi$, and $\rho_n^{(n)}(1)$ is $\varepsilon$ close to $\omega_n \otimes \psi$ as can be seen in \eqref{finiteCompOfEmb}. 
Let us also define 
\begin{equation} \label{tildeRho_iDef}
    \tilde \rho_i^{(n)} \equiv \Tr_{i+1,...,n}\omega_n \otimes \psi.
\end{equation}
With these definitions, a protocol with precision $1/\varepsilon$, archives $\|\rho_n^{(n)}(1)- \tilde \rho_n^{(n)}\|_1 \leq \varepsilon$.

\subsection{General bound}
In the following we derive a lower bound for the complexity of embezzlement that is independent of the catalyst's state, showing that it must diverge when $1/\varepsilon$ or the amount of embezzled entanglement grow to infinity. For this derivation, we use the methods of \cite{eisert2021entangling} that give a lower bound in terms of the state's change in entropy. 

Let us first introduce the property of \textit{small incremental entangling} \cite{bravyi2007upper, marien2016entanglement}: suppose there is a pure state $\chi \in \mathcal{H}_1 \otimes \mathcal{H}_2 \otimes \mathcal{H}_3 \otimes \mathcal{H}_4$ which at time $t$ evolves according to a Hamiltonian which couples only $\mathcal{H}_2 \otimes \mathcal{H}_3$. Denote the reduced density matrix of systems $1$ and $2$ by $\chi_{12} = \Tr_{3,4}(\chi)$. Small incremental entangling is the following inequality on the time derivative of $S(\chi_{12})$, the von Neumann entropy of $\chi_{12}$:
\begin{equation} \label{SIE}
    \frac{d S(\chi_{12}(t))}{dt}\leq g \log(D) \|H(t)\|,
\end{equation}
where $D = \min(\text{dim}(\mathcal{H}_2),\text{dim}(\mathcal{H}_3))$, and $g$ is a constant not larger than $22$.

The infinitesimal gate acting on the state at time $t$ is $e^{-i dt H(t)}$, which as $dt \to 0$ can be trotter decomposed to $\Pi_I e^{-i dt h_I(t)}$. We notice that at a given time, only the terms in $H$ that couple site $i$ and $i+1$, influence the change in the entropy of $\rho_i^{(n)}(t)$. For these terms, we can substitute $D=d$ in \eqref{SIE} to obtain,\footnote{When using \eqref{SIE} to derive \eqref{DeltaSbound}, we assume that $\mathcal{H}_2$ and $\mathcal{H}_3$ are the Hilbert spaces of sites $i$ and $i+1$, and $\mathcal{H}_1$ and $\mathcal{H}_4$ are the Hilbert spaces of all the sites before site $i$ and all the sites after site $i+1$ respectively. We note that if $d_e<d$, for $i=0$, the bound of \eqref{DeltaSbound} can be tightened by replace $d$ with $d_e$, but nevertheless it is valid in its current form.}
\begin{equation} \label{DeltaSbound}
\begin{split}
   |\Delta S(\rho_i^{(n)})| & \equiv  | S(\rho_{i}^{(n)}(1)) -  S(\rho_{i}^{(n)}(0))| 
    \\
    & = \left|\int_0^1 dt \frac{d}{dt}S(\rho_{i}^{(n)}(t))\right|
    \\
    & \leq \int_0^1 dt \left|\frac{d}{dt}S(\rho_{i}^{(n)}(t))\right|
    \\
    & \leq \int_0^1 dt g \log(d) \|\sum_{I_i} h_{I_i}(t)\|
    \\
    & \leq \int_0^1 dt g \log(d) \sum_{I_i} \| h_{I_i}(t)\|.
\end{split}
\end{equation}
where the triangle inequality has been repeatedly used, and the index $I_i$ runs over all operators which couple site $i$ and $i+1$. 
Noticing that $C(U)\geq \int_0^1 dt \sum_{i=0}^{n-1} \sum_{I_i} \| h_{I_i}(t)\|$ with equality if $h_I$ do not contain $1-local$ gates, and summing over the sites gives
\begin{equation} \label{C_nBound}
    C_n(\varepsilon) \geq \frac{1}{g \log(d)} \sum_{i=0}^{n-1} |\Delta S(\rho^{(n)}_i)|. 
\end{equation}
We are left to compute the sum but lack the needed details about $\rho^{(n)}_i(1)$. 
Instead, we shall compute the sum of 
\begin{equation} \label{defDFtil}
    |\Delta S(\tilde \rho_i^{(n)})| \equiv |S(\tilde \rho_i^{(n)}) - S(\rho_i^{(n)}(0))|,
\end{equation}  
and bound the difference with the sum of $|\Delta S(\rho_i^{(n)})|$. 

\textit{Fannes' inequality} \cite{fannes1973continuity} says that if two states $\rho$ and $\tilde \rho$ on a Hilbert space $\mathcal{H}$ are such that $\|\rho - \tilde \rho\|_1 = \varepsilon \leq 1/e$, then the difference in entropy is bounded by
\begin{equation} \label{Fannes}
    |S(\rho)-S(\tilde \rho)| \leq \varepsilon \log(\text{dim}(\mathcal{H})) - \varepsilon \log(\varepsilon).
\end{equation}

The trace norm of a partially traced operator is smaller or equal to the trace norm of the operator \cite{rastegin2012relations}, and therefore $\|\tilde \rho_i^{(n)} - \rho_i^{(n)}(1)\|_1 \leq \|\tilde \rho_n^{(n)} - \rho_n^{(n)}(1)\|_1 \leq \varepsilon$. 
This gives that 
\begin{equation} \label{DifSrhoStildrho}
\begin{split}
    \big| |\Delta S(\tilde \rho_i^{(n)})| - |\Delta S(\rho_i^{(n)})|\big| & \leq |S(\tilde \rho_i^{(n)}) - S(\rho_i^{(n)}(1))|
    \\
    & \leq \varepsilon \log(d_e d^{i})-\varepsilon \log(\varepsilon),
\end{split}
\end{equation}
where we have used the triangle inequality together with the definitions of $\Delta S(\rho_i^{(n)})$ and $\Delta S(\tilde \rho_i^{(n)})$ from \eqref{DeltaSbound} and \eqref{defDFtil} in the first line, and Fannes' inequality, \eqref{Fannes}, in the second. Therefore,
\begin{align}
     |\Delta S(\rho_i^{(n)})| \geq \max\left( |\Delta S(\tilde \rho_i^{(n)})|  - \varepsilon \log\left(\frac{d_e d^{i}}{\varepsilon}\right),0\right).
\end{align}
Let us denote by $\Delta S$ the amount of entanglement entropy that is being embezzled by the embezzling system in \eqref{finiteCompOfEmb}:
\begin{equation}
    \Delta S \equiv |S(\phi)-S(\psi)|.
\end{equation}
Because $\tilde \rho_i^{(n)}$ and $\rho_i^{(n)}(0)$ differ just by the factors of $\phi$ and $\psi$, we notice that $|\Delta S(\tilde \rho_i^{(n)})|= \Delta S$. This, together with \eqref{C_nBound}, gives the lower bound
\begin{align} \label{lowerbound}
   g \log(d) C_n(\varepsilon) \geq \sum_{i=0}^{n-1}\max\left( \Delta S  - \varepsilon \log\left(\frac{d_e d^{i}}{\varepsilon}\right),0\right).
\end{align}

There's a bound on the minimal possible value of $n$ for which it is possible to obtain a $1/\varepsilon$ precision, given by $\Delta S=|S(\tilde \rho_n^{(n)}) - S(\rho_n^{(n)}(1))|\leq \varepsilon \log(d_e d^{n}) -\varepsilon \log(\varepsilon)$. This means that we can express the bound as,
\begin{equation}
\label{lowerboundn_star}
\begin{split}
   g \log(d) C_n(\varepsilon) & \geq \sum_{i=0}^{\lfloor n_*\rfloor}\left(\Delta S  - \varepsilon \log\left(\frac{d_e d^{i}}{\varepsilon}\right)\right)
   \\
   & \geq \int_{0}^{ n_*} d i \left(\Delta S  - \varepsilon \log\left(\frac{d_e d^{i}}{\varepsilon}\right) \right),
\end{split}
\end{equation}
where $n_*$ is the value of $i$ for which the summand equals zero, and $\lfloor n_*\rfloor$ is the largest integer smaller than $n_*$. Notice that the bound is independent of $n$ and therefore also bounds $C(\varepsilon)$, as defined in \eqref{embezzlingComp}. 
This gives that
\begin{equation} \label{finalCbound}
\begin{split}
    C(\varepsilon)& \geq \frac{(\Delta S-\varepsilon \log\left(\frac{d_e}{\varepsilon}\right))^2}{2  \, g \, \log(d)^2 \, \varepsilon}
     \sim  \frac{\Delta S^2}{2 g \log(d)^2}\frac{1}{\varepsilon}.
\end{split}
\end{equation}
We obtained a lower bound for the complexity of embezzlement which diverges linearly with $\frac{1}{\varepsilon}$ and with $\Delta S^2$, demonstrating a physical obstruction for perfect embezzlement.
The bound in \eqref{finalCbound} is independent of the catalyst's state, and therefore embezzling families that enable its saturation are the most efficient.

As we show in Appendix \ref{SatOfComp}, the scaling with $\varepsilon$ in \eqref{finalCbound} can indeed be achieved using the embezzling family of van Dam and Hayden \cite{van2003universal}, provided the reduced density matrix $\omega_n$ is diagonal in a suitable basis. This may not be surprising, as the van Dam--Hayden family approximately achieves the minimal number of sites required for embezzlement with precision $1/\varepsilon$ \cite{van2003universal}. However, saturating the bound demands a specifically fine-tuned embezzling family.

We recall that a primary motivation for this work is the claim that physical systems, such as relativistic quantum field theories and certain lattice models in the thermodynamic limit, allow embezzlement to arbitrary precision \cite{van2024critical, van2024embezzlement, van2024embezzling}. Accordingly, we now turn to study the complexity of embezzlement when the catalyst belongs to these systems.

\subsection{Embezzling from a critical system.} \label{criticalSec}
Let us consider the catalyst to be a $1d$ critical system in its vacuum state, where universal properties allow for analytic control. One may expect the complexity of embezzlement to grow more quickly in such systems if the relevant embezzling families have a slower growth of the optimal precision with the system size. Since the bound of \eqref{lowerboundn_star} does not depend on the specific nature of the catalyst, we now derive a different lower bound. The starting point is the lower bound, derived in Appendix \ref{SchattenNormBound} on the complexity of state preparation that depends on the Schatten $p$-norms, $\|*\|_p$ of partitions of the state. Using \eqref{CboundAppen} for the embezzlement protocol gives,
\begin{equation} \label{CboundSchatten}
    2 C_n(\varepsilon)\geq \sum_{i=0}^{n-1}\Delta \|\rho_i^{(n)}\|_{p(i)},
\end{equation}
where $C_n(\varepsilon)$ is defined in \eqref{finiteCompOfEmb}, $p(i) \in [1,\infty]$, and  $\Delta \|\rho_i^{(n)}\|_p  \equiv  \Big| \|\rho_{i}^{(n)}(1)\|_p -  \|\rho_{i}^{(n)}(0)\|_p\Big|$.
We remind that using the notation defined around \eqref{tildeRho_iDef}, the state $\tilde \rho_n^{(n)}$ satisfies
$\|\tilde \rho^{(n)}_n-\rho^{(n)}_n(1)\|_1 \leq \varepsilon$.
Instead of computing $\Delta \|\rho_i^{(n)}\|_p$, we shall evaluate a lower bound on the complexity in \eqref{CboundSchatten}, using the sum over 
\begin{equation}
   \Delta \|\tilde \rho_i^{(n)}\|_p  \equiv  \Big| \|\tilde \rho_{i}^{(n)}\|_p -  \|\rho_{i}^{(n)}(0)\|_p\Big|.
\end{equation}
The difference between $\Delta \|\rho_i^{(n)}\|_p$ and $\Delta \|\tilde \rho_i^{(n)}\|_p$ can be bounded by,
\begin{equation}\label{boundnAppen}
\begin{split}
    \Big|\Delta \|\rho_i^{(n)}\|_p  -  \Delta \|\tilde \rho_i^{(n)}\|_p  \Big|
    & \leq \Big| \|\rho_{i}^{(n)}(1)\|_p  -  \|\tilde \rho_{i}^{(n)}\|_p \Big|
    \\
    & \leq \|\rho_{i}^{(n)}(1) -  \tilde \rho_{i}^{(n)}\|_p
    \\
    & \leq \|\rho_{i}^{(n)}(1) -  \tilde \rho_{i}^{(n)}\|_1 
    \\
    & \leq \varepsilon,
    \end{split}
\end{equation}
where we have used the triangle inequality in the first and second lines, and the relations of \eqref{monoNorms} in the third and fourth lines. Using this to further lower bound \eqref{CboundSchatten}, gives
\begin{align}\label{lowerboundSchatten}
    2 C_n(\varepsilon) \geq \sum_{i=0}^{n-1} \max\left(\Delta \|\tilde \rho_i^{(n)}\|_{p(i)} - \varepsilon,0\right).
\end{align}
The above lower bound is valid for any $p(i)$. Therefore, the tightest bound will be given by a choice of $p(i)$ that maximizes $\Delta \|\tilde \rho_i^{(n)}\|_{p(i)}$.

We now assume that $\phi$ in~\eqref{rhoiDef} is pure, so that 
$\|\phi\|_p = 1$.  We also take the state $\psi$ in~\eqref{tildeRho_iDef} to be maximally mixed (rank $d_e$ with equal eigenvalues), implying 
$\|\psi\|_p = d_e^{1/p - 1}$.  Physically, this corresponds to embezzling a maximally entangled pair 
of $d_e$ level systems.  Under these assumptions we find
\begin{equation}\label{deltaTilRho}
  \Delta \|\tilde{\rho}_i^{(n)}\|_{p(i)}
  \;=\;
  \bigl(1 - d_e^{\,\frac{1}{p(i)} - 1}\bigr)\,
  \|\rho_i^{(n)}(0)\|_{p(i)},
\end{equation}
where $\|\rho_i^{(n)}(0)\|_{p(i)} = \|\Tr_{i+1,\dots,n}\omega_n\|_{p(i)}$ since $\|\phi\|_p = 1$.

In the case of a $1d$ critical system, $\omega_n$ can represent, for example, 
the reduced density matrix of $n$ sites in a critical spin chain, or the reduced density matrix 
of a subregion of size $l$ in a conformal field theory with a UV cutoff 
$\epsilon_{\mathrm{UV}} = l / n$. In the latter scenario, it is worth clarifying that the UV cutoff provides $n$ modes, and one may trace out all but $i$ of them. 
These modes could be spatial, as in a lattice discretization, or in momentum space (see, \eg \cite{schwartzman2021entanglement}).

For large enough subregions in such systems, the Schatten $p$ norm takes the form 
\cite{calabrese2004entanglement, calabrese2008entanglement,calabrese2010parity}
\[
\|\rho_i^{(n)}(0)\|_p 
\;=\; (c_p)^{1/p} \,\exp\biggl[-\,b\!\Bigl(1 - \tfrac{1}{p^2}\Bigr)\biggr],
\]
where $b = \frac{c}{6}\,\ln\!\bigl(L_{\mathrm{eff}}(i,n)\bigr)$, $c$ is the central charge, 
and $c_p$ is a non-universal constant. The length scale $L_{\mathrm{eff}}$ depends on the setting: 
it equals $i$ for a subregion of an infinitely long system, or $\frac{2n}{\pi}\,\sin\!\bigl(\frac{\pi i}{2n}\bigr)$ 
for a system of finite length with $2n$ sites and periodic boundary conditions. 
In what follows, we focus on the former, where 
$b(i) = \frac{c}{6}\,\ln(i)$.

It was shown that the assumption of constant $c_p$ provides a good approximation for the behavior of critical spin systems \cite{calabrese2008entanglement, pollmann2010entanglement}. 
For the models considered in \cite{pollmann2010entanglement}, its value deviates from 1 by at most $10 \% $. From here on we shall assume that $c_p=1$. 
If a particular model exhibits a $p$–dependent coefficient $c_p$ that is nonetheless independent 
of the block size $i$, one can simply replace the factor $(c_p)^{1/p}$ by 
$\min_{p \ge 1} (c_p)^{1/p}$ in all subsequent expressions.\footnote{
The scaling of our bound is governed by the behaviour of 
$\Delta\|\tilde{\rho}_i^{(n)}\|_{p(i)}$ near $p(i)=1$.  
If $c_p\neq 1$, the first‑order correction involves 
$\partial_p c_p\,\bigl|_{p=1}$, which is the non‑universal constant term that appears in 
the entanglement entropy.}

To optimize the lower bound in \eqref{lowerboundSchatten}, we begin by choosing a suitable 
function $p(i)$. We focus on the summand in 
\eqref{lowerboundSchatten} at large $i$, since this regime dominates the bound as $\varepsilon$ 
decreases. In this limit, $b$ is large, and the maximum of \eqref{deltaTilRho} occurs near 
$p-1 \,\sim\, \tfrac{1}{2b}$. Thus, for simplicity, we set
\begin{equation}\label{pOfi}
  p(i) \;=\;\Bigl(1 \;-\;\tfrac{1}{2\,b(i)}\Bigr)^{-1},
  \quad
  \text{for } b(i) > \tfrac12.
\end{equation}
With this choice, we find that
\begin{equation}\label{deltaRhoHere}
  \Delta \|\tilde{\rho}_i^{(n)}\|_p 
  \;=\; 
  e^{\tfrac{1}{4b} - 1}\,
  \Bigl(1 \;-\; d_e^{-\tfrac{1}{2b}}\Bigr).
\end{equation}
Because $\|\rho_n^{(n)}(0)\|_p= \|\rho_n^{(n)}(1)\|_p$, setting $i = n$ in \eqref{boundnAppen} yields
\begin{equation}\label{OneMoreBound}
    \Delta \|\tilde{\rho}_n^{(n)}\|_p 
    \leq
    \varepsilon.
\end{equation}
This implies that for any catalyst state permitting embezzlement with precision $1/\varepsilon$, 
one can evaluate the bound \eqref{lowerboundSchatten} with \eqref{deltaRhoHere} inserted, by summing up to the index where 
the summand vanishes, as it will always be smaller or equal to $n$. In addition, as this lower bound is independent of $n$, it also bounds $C(\varepsilon)$.
Bounding the sum of the monotonic decreasing function with an integral gives
\begin{equation}\label{criticalLower}
\begin{split}
    2 C(\varepsilon) & \geq \int_{1}^{i_*} di \Delta \|\tilde \rho_i^{(n)}\|_{p(i)}- \varepsilon 
    \\
    & \geq \int_{i_{1/2}}^{i_*}di e^{\frac{1}{4 b(i)} -1}\left(1-e^{-\frac{\log(d_e)}{2 b(i)}}\right) - \varepsilon
    \\
    & \geq \int_{i_{1/2}}^{i_*}di e^{ -1}\left(1-e^{-\frac{\log(d_e)}{2 b(i)}}\right) - \varepsilon
    \\
    & \geq \int_{i_{1/2}}^{i_*}di \frac{1}{e}\left(\frac{1}{1+\frac{2b(i)}{\log(d_e)}}\right) - \varepsilon
    \\
    & =    \frac{\log({d_e}^{\frac{3}{c \, e}})}{{d_e}^{3/c} }   \, \left( \text{li}({d_e}^{\frac{3}{c \, e \, \varepsilon}}) - \text{li}({d_e}^{\frac{3}{c}} i_{1/2})\right)
    \\
    & \ \ \ \ \ \ \ \ \ \ \ \ \ \ \ \ \ \ \ \ \ \ \ \ \ \ \ \ \ \ \ \ \ \ \ \ \ \   -(i_*-i_{1/2})\varepsilon
    \\
    & \sim \frac{\varepsilon^2 }{ \log({d_e}^{\frac{3}{c \, e}})} {d_e}^{\frac{3-3 e \varepsilon}{c \, e\, \varepsilon}} 
    \\
    & = \frac{c \, e \,\varepsilon^2 }{ 3 \Delta S} {e}^{\frac{\Delta S}{c \, e\, \varepsilon} (3-3 e \varepsilon)} 
\end{split}
\end{equation}
where $i_*$ is the value of $i$ for which the integrand vanishes, and its value at each line changes accordingly. $i_{1/2} = e^{3/c}$, is the value at which $b(i_{1/2})=1/2$, and li($\cdot$) is the logarithmic integral function. The second inequality is given because the integrand is positive and the interval of integration, $[1,i_*]$, contains the interval $[i_{1/2},i_*]$. In the third inequality we notice that $e^{\frac{1}{4 b(i)}}$ is greater than $1$, and in the fourth we used the fact that $1-e^{-1/x}\geq\frac{1}{1+x}$ for $x>0$.  We find that for $1d$ critical catalysts, the complexity grows with the precision faster than a (polynomially diminished) exponential, making it a highly inefficient resource for embezzlement. 

Finally, we would like to point out that \eqref{OneMoreBound} gives a lower bound for the number of sites, $n_{crit}$, or the UV cutoff in the critical system, needed to perform the embezzlement with precision $1/\varepsilon$. With the same choice for $p(i)$ as in \eqref{pOfi}, and using similar reasoning to the inequalities of \eqref{criticalLower} for lower bounding $\Delta \|\tilde \rho_n^{(n)}\|_p $, we obtain\footnote{We leave for the future the computation of $n_{crit}$'s exact value using the entanglement spectra of $1d$ critical systems found in \cite{calabrese2008entanglement}. }
\begin{equation}\label{minimalSites}
    n_{crit} = \frac{l}{\epsilon_{UV}} \geq {d_e}^{\frac{3-3 e \varepsilon}{c \, e\, \varepsilon}} ={e}^{\frac{\Delta S}{c \, e\, \varepsilon} (3-3 e \varepsilon)} .
\end{equation}

\subsection{Other circuit cost models }\label{Other}
So far we have restricted the discussion for control Hamiltonians which couple nearest neighbor sites. We now describe how variations of the circuit cost model can affect the lower bound for the complexity of embezzlement, obtained in \eqref{finalCbound} and \eqref{criticalLower}.

If the coupling allowed is between $k>2$ nearest neighbors, there will be a slight modification that will not change the functional dependence on $\varepsilon$ and $\Delta S$. When using \eqref{SIE} to estimate an upper bound for $|\Delta S(\rho_i^{(n)})|$, if $H(t)$ is a $k$ nearest neighbor interaction term, the value $D$ takes depends on where these $k$ sites are with respect to site $i$. However, the maximal value $D$ can have in this case is $d^{\lfloor k/2 \rfloor}$, and therefore in \eqref{DeltaSbound}, a correct bound can be obtained by changing $\log(d)$ to $\log(d^{\lfloor k/2 \rfloor})$. In addition, the sum $\sum_{i=0}^{n-1}|\Delta S(\rho_i^{(n)})|$ will over-count $\|h_I\|$ terms that affect the entropy of multiple $\rho_i^{(n)}$s. For example, in a spin chain, the term $\sigma^1_a \otimes \sigma^2_b \otimes \sigma^3_c$ where the superscript denotes the site, can enter both in the $|\Delta S(\rho_1^{(n)})|$ and $|\Delta S(\rho_2^{(n)})|$ terms. This will modify \eqref{C_nBound} such that there is an overall factor on the right-hand side that will account for the over-counting and the change in the $k$ - locality of the interactions (alternatively, one can alter the sum in \eqref{C_nBound} to $\sum_{i=0}^{ i = \left \lfloor{n/(k-1)}\right \rfloor} |\Delta S(\rho_{(k-1) i}^{(n)})|$, which will cancel the over-counting).  Nevertheless, this factor is independent of $n$ and will not change the $\frac{1}{\varepsilon}$ divergence seen in \eqref{finalCbound}. An analogous analysis also applies to the derivation of \eqref{CboundSchatten}, changing the result in \eqref{criticalLower} by an overall factor that is independent of $\varepsilon$ and $\Delta S$. Similar arguments can show a power-law divergence with $1/\varepsilon$ and $\Delta S$ for the embezzling complexity lower bound in higher dimensional lattices with $k$-geometrically local control Hamiltonians.

In addition, consider the case when the control Hamiltonian is allowed to have $2$-local interactions between any two sites. In that case, the sum in \eqref{DeltaSbound} will be over all $\|h_I\|$s that couple sites $k$ and $p$, with $k\leq i$ and $p>i$. Summing \eqref{DeltaSbound} over all sites will give 
\begin{equation}
    \frac{1}{c \log(d)} \sum_{i=0}^{n-1} |\Delta S(\rho_i)|\leq \sum_I q_I \|h_I\|,
\end{equation}
where $q_I$ equals the distance between the sites. Therefore, \eqref{finalCbound} will bound the complexity of an embezzling protocol with $2$-local control Hamiltonians (not necessarily nearest neighbors), where the cost functional is $\int_0^1 dt \sum_I p_I \|h_I\|$ and the penalty for a two-site coupling, $p_I$, is at least as large as the distance between the sites.

As a final variant, we consider a coarse-grained embezzling protocol in which the one performing the protocol can manipulate $n$ sites of the catalyst, but at the final stage, they are only interested in the $m<n$ first sites. More explicitly, we consider 
\begin{equation}\label{variant}
    \begin{split}
        C_{n,m}(\varepsilon) & = \min_{U}  C(U) 
        \\
        & \text{s.t} \ \ \| \Tr_{m+1,...,n}(U \omega_n \otimes \phi U^\dagger - \omega_n \otimes  \psi)\|_1 < \varepsilon.
    \end{split}
\end{equation}
In this case, the difference in \eqref{DifSrhoStildrho}, can only be bounded for $i\leq m$. However, in \eqref{lowerbound}, summing up to $m-1$ instead of $n-1$ will still give a valid lower bound, which will have a maximal value when $\varepsilon \to 0$, of $\frac{m \Delta S}{c \log(d)}$. This bound diverges only with $\Delta S$, with no $\varepsilon$ dependence.

\section{Discussion and future developments}

In this work, we studied lower bounds on the circuit complexity of entanglement embezzlement.  
We derived a general bound that grows at least polynomially with the embezzlement precision or the amount of embezzled entanglement, and diverges in the limit of perfect precision or infinite entanglement.  As the general bound is independent of the catalyst's details, any protocol saturating it is the most efficient one.
Thus, we argued that the counterintuitive protocol of embezzlement is physically constrained by circuit complexity.

Circuit complexity in quantum field theories remains under active research and is not yet sufficiently developed to be computed for most unitary operations in the theory. Our results, however, apply equally to catalyst systems that can be viewed as regularized quantum fields and are independent of the regularization cutoff. In this way, we have provided a well-defined model of circuit complexity for quantum fields and demonstrated its divergence for perfect embezzling protocols. As a concrete example, we studied the protocol using a $1d$ critical system and showed that the complexity grows superpolynomially (at least an exponential tempered by a polynomial factor) with both precision and the amount of embezzled entanglement.

One development we leave for the future is to bound the embezzling complexity for Gaussian unitaries. It is possible that perfect entanglement embezzling from a free relativistic quantum field can be achieved with Gaussian operations alone. Using similar techniques, we plan to show a divergent lower bound on this complexity without constructing the explicit unitary protocol.

As the diverging complexity of perfect embezzlement agrees with physical intuition, it is interesting to ask a converse question. Suppose that any physical complexity measure must diverge in the limit of perfect entangling embezzling. What constraints does this impose on the possible complexity models, and how do they scale with the system's dimension? Specifically for Nielsen complexity, what types of norms are physical in this sense?

Finally, we hope to explore a potential relation to holography. At the heart of the black hole information puzzle is the seemingly large entropy of the Hawking radiation after the Page time \cite{page1993information}, where to conclude the state evolved unitarily, one needs to perform a high complexity decoding \cite{harlow2013quantum}. It has been argued that for two boundary conformal field theories to describe a two-sided black hole with a sharp horizon, there must be an emergent type III$_1$ algebra of operators in each boundary \cite{leutheusser2023causal}. Therefore, in such setting, it should be possible to perform an embezzling protocol where we expect that the precision or the amount of embezzled entanglement possible will be controlled by $N$, the number characterizing the degrees of freedom in the conformal field theory. 
One can also consider a one-sided protocol such that the state dual to a two-sided black hole is taken to a new pure state that is close on one side to a state with higher entropy while on the other side it remains unchanged. This means that restricted to each side, the new pure state is arbitrary close to a target state that is mixed globally. However without a restriction to one side, the trace distance between the two states, one being pure and the other mixed, has to be large. This suggests that if bulk geometrical duals exist for the two states, they will differ significantly only inside the black hole interior, or on its horizon. In addition one can study the protocol in Rindler-AdS/CFT, where the precision and entanglement of embezzlement between the Rindler wedges will be controlled by the UV cutoff, in addition to $N$. A recent work studies in holography a \emph{spoofing} phenomenon in which two states have significantly different entropy but one needs a complex protocol to distinguish between them \cite{engelhardt2024spoofing}. It would be interesting to understand how the embezzlement protocol and its complexity manifest in the bulk dual, what are its limitations for finite $N$, and what connections are there to spoofing entanglement in holography.

\begin{acknowledgments}
The author would like to greatly thank Shira Chapman for her guidance, support, and detailed comments on the manuscript. The author also thanks Michael Chapman, Lauritz van Luijk, Alexander Stottmeister, Henrik Wilming, Yaron Oz, Lorenzo Di Pietro, Dami\'an A. Galante, and Saskia Demulder, for helpful discussions. This work is supported by the Israel Science Foundation
(grant No. 1417/21), by the German Research Foundation through a German-Israeli Project Cooperation (DIP) grant “Holography and the Swampland”, by Carole and Marcus Weinstein through the BGU Presidential Faculty Recruitment Fund, by the ISF Center of Excellence for theoretical high energy physics and by the ERC starting Grant dSHologQI (project number 101117338).
\end{acknowledgments}

\appendix

\section{Saturation of the complexity bound} \label{SatOfComp}
In this section, we show that the scaling with $\varepsilon$ in \eqref{finalCbound} can be achieved with a protocol done with the embezzling family of van Dam and Hayden \cite{van2003universal}, given that the reduced density matrix $\omega_n$ is diagonal in a convenient basis. 
For this embezzling family, as shown in \cite{van2003universal}, the optimal protocol yields $\varepsilon \sim \frac{1}{n}$. Consider a system of qubits, where the embezzling system is initially in a state $\phi =\ket{0}\bra{0}$, 
and 
\begin{equation}
\omega_n \otimes \phi =  \lambda_1 \ket{00...0}\bra{00...0}+\lambda_2 \ket{10...0}\bra{10...0}+...\, ,
\end{equation}
where the spectrum is in decreasing order according to the binary representation of the computational basis. Suppose we wish the embezzle a maximally entangled qubit pair, such that $\psi = \frac{1}{2}\ket{0}\bra{0}+\frac{1}{2} \ket{1}\bra{1}$. In this case,
\begin{equation}
\omega_n \otimes \psi =  \frac{\lambda_1}{2}\left(\ket{00...0}\bra{00...0}+\ket{00...1}\bra{00...1}\right) + ... \, .
\end{equation} 
The protocol for embezzlement which minimizes $\| U \omega_n \otimes \phi U^\dagger - \omega_n \otimes  \psi\|_1$ will be realized by a unitary that rearranges $\omega_n \otimes \phi$ in the basis in which $\omega_n \otimes \psi$ is in a decreasing order \cite{powers1967representations}, \ie 
\begin{equation}
U \omega_n \otimes \phi U^\dagger =  \lambda_1 \ket{00...0}\bra{00...0}+\lambda_2 \ket{00...1}\bra{00...1}+... \, .
\end{equation}
This can be done with a unitary composed out of $n$ consecutive nearest neighbor swaps, realizing a bitwise circular shift. Therefore, the cost of this unitary grows with $n$, and as the minimal $n$ needed for obtaining $\varepsilon$ grows as $\frac{1}{\varepsilon}$, we obtain $C(\varepsilon) \sim \frac{1}{\varepsilon}$.

\section{Schatten norm complexity bound}\label{SchattenNormBound}

Here we present lower bounds for the circuit complexity in terms of the change in the Schatten $p$ norms of the partial traces of the initial and final state. These can be used to obtain a lower bound on the complexity of embezzlement but could be of independent interest, complementing the entropy lower bounds of \cite{eisert2021entangling}. Unlike the results of \cite{eisert2021entangling}, our bound does not depend on the dimension of the qudits on which the circuit acts.

Suppose that $\rho^{(n)}(0)$ is a state of a qudit chain with $n+1$ sites where the local Hilbert space dimension of each site is $d$. We are interested in bounding the minimal cost of all circuits that transform $\rho^{(n)}(0)$ to $\rho^{(n)}(1)$. The circuits are generated by a $2$-geometrically local Hamiltonian, as in \eqref{controlH} and the cost is computed by \eqref{costNorm}. Let us denote this quantity by
\begin{equation}
\begin{split}
    C( & \rho^{(n)}(0),\rho^{(n)}(1)) 
    \\
    & = \min_{\lbrace Y^I : \, U(0)=\mathbf{1}, \, U(1)\rho^{(n)}(0)U(1)^\dagger=\rho^{(n)}(1) \rbrace} \text{cost}(U(t))\, .
\end{split}
\end{equation}

Let us also define $\rho^{(n)}(t) = U(t)\rho^{(n)} U(t)^\dagger$, the state evolved along the circuit which gives the minimal cost, and $\rho^{(n)}_i(t) =  \Tr_{i+1,...,n}\rho^{(n)}(t)$, the reduced density matrix of the first $i+1$ sites (where the counting starts from site number $0$). We define $h^i = \sum_{I_i} h_I$ as the sum of all terms in the Hamiltonian which couple the sites $i$ and $i+1$. The Schatten $p$ norm of a linear bounded operator between two Hilbert spaces $A$, is defined as,
\begin{equation}
    \|A\|_p = \left(\Tr(|A|^p)\right)^{1/p},
\end{equation}
where $p\geq 1$. The norms are not influenced by local unitary operations, and therefore the time derivative of $\|\rho_i^{(n)}(t)\|_p$ will only be influenced by the $h^i$ term of the control Hamiltonian. This gives the following lower bound: 
\begin{equation}
\begin{split}
   & \Delta \|\rho_i^{(n)}\|_p  \equiv  \Big| \|\rho_{i}^{(n)}(1)\|_p -  \|\rho_{i}^{(n)}(0)\|_p\Big|
    \\
    & = \Big|\int_0^1 dt \frac{d}{dt}\|\rho_{i}^{(n)}(t)\|_p\Big|
    \\
    & \leq \int_0^1 dt \bigg|\frac{\|\rho_{i}^{(n)}(t)+i dt\Tr_{\bar i}\left( [h^i,\rho^{(n)}(t)]\right)\|_p-\|\rho_i^{(n)}(t)\|_p}{dt}\bigg|
    \\
    & \leq \int_0^1 dt \|\Tr_{\bar i}\left( [h^i,\rho^{(n)}(t)]\right)\|_p,
\end{split}
\end{equation}
where $\Tr_{\bar i} \equiv \Tr_{i+1,...,n}$, and the inequalities were obtained by using the triangle inequality. 
Recall that the Schatten norms are monotonic with respect to $p$, for $p \geq 1$, and in addition, the trace norm (which equals the Schatten 1-norm), is monotonic under a partial trace \cite{rastegin2012relations}, \ie the norms satisfy
\begin{equation} \label{monoNorms}
    \|*\|_p \leq \|*\|_1 \ \ \ ; \ \ \ \|\Tr_{\bar i}(*)\|_1 \leq \|*\|_1.
\end{equation}
Using this observation, we obtain,
\begin{equation}
\begin{split}
   \Delta \|\rho_i^{(n)}\|_p 
    & \leq \int_0^1 dt  \|\Tr_{\bar i}\left( [h^i,\rho^{(n)}(t)]\right)\|_1
    \\
    & \leq \int_0^1 dt  \| [h^i,\rho^{(n)}(t)]\|_1
    \\
    & \leq \int_0^1 dt  \| h^i \rho^{(n)}(t)\|_1 + \| \rho^{(n)}(t) h^i\|_1
    \\
    & \leq \int_0^1 dt  2\| h^i\| \|\rho^{(n)}(t)\|_1 = 2\int_0^1 dt \| h^i\|,
\end{split}
\end{equation}
where the triangle inequality has been used in the third inequality, and H\"{o}lder's inequality together with the fact that any state has unit trace norm gave the final line. Notice that the above is valid for any choice of $p$.

Summing over all sites, with the possibility of different $p \geq 1$ for each of them, yields a lower bound for the minimal cost:
\begin{equation} \label{CboundAppen}
    C(\rho^{(n)}(0),\rho^{(n)}(1))\geq \frac{1}{2}\sum_{i=0}^{n-1}\Delta \|\rho_i^{(n)}\|_{p(i)}.
\end{equation}
We note that this lower bound is upper bounded by $n$, and therefore cannot show a growth stronger than linear with the number of sites.

\bibliography{Bibliography}
\bibliographystyle{unsrt}

\end{document}